\documentstyle{article}
\begin{document}
\title{\Large \bf Effect of Level Statistics on Local Magnetism
                 in Nanoscale Metallic Grains \\}
\author{\large Liang-Jian Zou and Zheng Qing-Qi$^{1}$  \\
{\it Institute of Solid State Physics, Academia Sinica,
      P.O.Box 1129, Hefei 230031, China} \\
{\it $^{1}$ Hefei Center of Advanced Studies, University of Science and 
      Technology of China, Hefei 230031, China.} \\}
\date{ }
\maketitle
\large

\begin{center}
{\bf Abstract} \\
\end{center}

    Effect of level statistics on local electronic states and local magnetism
in nanoscale metallic grains with transition-metal impurity in the ballistic
regime is studied. It is shown that the mean occupation of local
electron and the local magnetic moment in nanoscale metallic grains with
odd conduction-electrons are larger than those with even conduction-electrons.
The effect of even-odd parity on the condition for the occurrence
of local magnetic moment is also discussed, it is found that the critical
value, $\rho_{d}(0) U_{c}$, for the formation of local moment in
nanoscale metallic grains is much smaller than that in bulks. The dependences
of the local spin susceptibility on size and the Coulomb interaction are
obtained. These results show that the level statistics plays an important role
for the local magnetism, it distinguishes
the properties of nanoscale metallic grains from those of clusters and bulks.

\vspace{2cm}

\large
\noindent{\em PACS numbers:} 75.20.Hr, 36.40.Cg, 73.23.Ps, 71.24.+q \\

\noindent{\it Keywords: Level Statistics, Local electronic states, Local
magnetic moment, Nanoscale metallic grains, Even-odd parity. }

\newpage
\large

\noindent {\bf 1. INTRODUCTION} \\

   Nanoscale metallic grain, varying from a few nanometers to several hundred
nanometers in size is complicated. It is neither few-body system as small
cluster, in which the localization character of electronic states is dominant,
nor an infinite system with translation invariance approximately as bulk, in
which the band character of electronic states is predominant. When the size of
grain, L,
is comparable with the mean free path of the conduction electrons,
two mesoscopic effects, the phase interference and the level
statistics, play important roles for the low-temperature properties of
nanoscale metallic grain. First, the average separation of the energy level,
\( \delta_{L} \approx E_{F}/N \),  is about 0.01 to 1K, here \( E_{F} \) is
the Fermi energy and N the number of conduction-electrons,
hence the quantum size effect including the size-dependence of energy levels
and the level repulsion should also be considered at low temperature.
Second, the grain size, L, may
be shorter than the phase coherence length \( L_{\phi} \), so the
quantum interference of the phases of electron wavefunctions plays an important
role. These two effects describe the quantum nature of many-electron
interaction and single-electron properties in nanoscale and mesoscopic systems.
These effects distinguish the physical properties of small metallic grains
from those of bulks or clusters.

        The magnetic properties of small clusters and disorder systems have
attracted many authors' attention$^{1-4}$ in the past few
years. The effect of size-dependence of the energy level on magnetism in small
perfect clusters has been discussed by some authors (for example,
see Refs. 1-2), the effect of the level statistics is not taken
into account in such small perfect clusters with regular boundary since the
effect of the statistical fluctuation of the levels on the magnetic properties
is not as important as that of the size effect. When the grain size
increases to a few nanometers and the disorder is taken into account, these
mesoscopic effects take their positions. In the same time, an elastic
scattering lifetime, $\tau$, should be introduced for the propagation of
conduction electrons in diffusive grain, the local magnetism will be affected
by the disorder of metallic grains. This point
is also not considered in the previous studies.
Recently, the influences of the level statistics arising from the quantum size
effect $^{3}$ and the mesoscopic fluctuation arising from the quantum
interference
effect $^{4}$ on the spin magnetism of the conduction electrons in disorder
mesoscopic metallic grains are studied. However, the local magnetism in
nanoscale disordered metallic grains has not yet been studied systematically.
Understanding this problem can provide insight the evolution of the local
magnetism from clusters, grains to bulks at low temperature.

  In the problem of local magnetism in ballistic grains, the quantum size
effect of the levels plays a more important role than the quantum
interference effect does. This effect manifests itself in the size-dependence
of energy levels and the statistical fluctuation of energy levels. And the
even-odd parity of the metallic
grain modifies the electronic states, hence the local magnetism. In the present
study, we find that the formation of local magnetic moment in
nanoscale metallic grains with odd conduction-electrons is more easier
than that with even conduction-electrons. We arouse an argument that
there still exists a critical condition for the occurrence of local magnetic
moment, it exhibits strong size dependence, however.

   In this paper, we concentrate on the ballistic metallic grains with dilute
transition metal impurities, the separation between impurities is much
larger than the magnetic scattering length  $L_{s}$, so that the
impurities behave as
isolated. The rest of this paper is organized as following: in Sec.2,
the formalism of the local magnetism in the ballistic grains is described;
the results and discussions are given in Sec. 3 and the conclusion is shown
in Sec. 4. \\

\noindent {\bf 2. MODEL and FORMALISM} \\

  Magnetic impurity, such as transition-metal atom ( Fe, Co, Ni, Mn,etc.),
dissolved in nonmagnetic metallic host can be described by the Anderson
impurity model $^{5}$. The Anderson model
provides an essential description for the interaction between conduction
electrons and the local electron. It explains the mechanism of the formation of
magnetic moment in bulk metal and plays an important role in our understanding
on the existence of local magnetic moment (LMM) in infinite system. This model
is believed to be still valid in describing the physical process in metallic
grains, and expressed in the following:
\begin{equation}
    H = \sum_{k\sigma}\epsilon_{k\sigma}
c^{\dag}_{k\sigma}c_{k\sigma} +
 \sum_{\sigma}[\epsilon_{d\sigma} d^{\dag}_{\sigma}d_{\sigma}
  + \frac{U}{2} n_{\sigma}n_{\bar{\sigma}}] +
   \sum_{k\sigma}[V_{k} c^{\dag}_{k\sigma}d_{\sigma}+ h.c.]
\end{equation}
where c$^{\dag}_{k\sigma}$ creates the
conduction (s) electron with state ${\bf k}$ and spin ${\it \sigma}$, and
$\epsilon_{k\sigma}=\epsilon_{k}-\sigma \mu_{B}B$, denotes the free-electron
dispersion under magnetic field B; d$^{\dag}_{\sigma}$ creates
a {\it d}-electron of the transition-metal impurity with the energy level
$\epsilon_{d\sigma}$=$\epsilon_{d}-g\sigma \mu_{B}B$, and $U$ represents the
on-site Coulomb interaction of the d-electrons; the hybridization matrix
element between the conduction- and the d- electrons is $V_{k}$.
In the absence of the hybridization interaction, the retarded propagator of the
s-electron is: \begin{equation}
   G_{0,\sigma}(z+i\eta)= \frac{1}{z-\hat{H_{s}}+i \eta}
\end{equation}
and that of d-electron is:
\begin{equation}
   D_{0,\sigma}(z+i\eta)= \frac{1}{z-\hat{H_{d}}+i \eta}
\end{equation}
In the presence of disorder, the conduction electron in metallic grains is
scattered by the nonmagnetic disorder potential  {\it v}(${\bf r}$), it is
assumed that the potential has vanish average, and
their correlation is  $\delta$-like, i.e., $\ <$ ${\it v}$ (${\bf r}$) $\ >$
=0 and $\ <$ ${\it v}$ (${\bf r}$) ${\it v}$ (${\bf r}^{'}$) $\ >$=
$\ <$ ${\it v}^{2}$ $\ >$ $\delta$(${\bf r-r'})$. Assuming the metallic grain
is weak disordered, then
the problem can be described in the framework of the perturbation theory.
According to Altshuler et al. $^{6,7}$, after the average to the disorder
configuration, the disorder effect may lead to a finite lifetime of the
quasiparticles $\tau$,  1/$\tau$=$\pi \rho_{0}$$\ < {\it v}^{2} \ >$, where
$\rho_{0}$ is the density of states for the conduction electrons near the Fermi
energy, then the propagator of the conduction electrons is rewritten
after the disorder average to Eq.(2):
\begin{equation}
   G_{0,\sigma}(z) = 1/(z-\epsilon_{k}+i \frac{\hbar}{2\tau})
\end{equation}
The propagator of the localized d-electrons is not affected by the
nonmagnetic disorder:
\begin{equation}
   D_{0,\sigma}(z) = \frac{1-<n_{\bar{\sigma}}>}{z-\epsilon_{d}}
                   + \frac{<n_{\bar{\sigma}}>}{z-\epsilon_{d}-U}
\end{equation}
According to Rice $^{8}$, the physical meaning of 1/$\tau$ in (4) is that the
phase correlation among the plane waves for conduction electrons is lost in a
distance ${\it l=v_{F}}$$\tau$, $v_{F}$ is the Fermi velocity. As a result,
the averaged propagator is exponentially damped in the distance {\it l},
this is the so-called weak localization resulting from the back scattering
of electron wavefunction by the disorder potential interfering with itself.
In the present paper, we concentrate on the metallic grain in the ballistic
regime, i.e., the diffusive length $l$ is much larger than
the sample size L, so the inverse of the lifetime, 1/$\tau$, can be regarded as
vanish, the propagator in
Eq.(4) becomes the usual form of free conduction-, or ${\it s}$-, electrons.

  The presence of the hybridization interaction will lead to self-energy
corrections to the propagators of the ${\it s-}$electrons and ${\it
d-}$electrons.  By the perturbation theory, it can be shown
that in the self-consistent mean-field approximation, the propagator of the
d-electron is:
\begin{equation}
    D_{\sigma}(\omega) =D_{0,\sigma}(\omega)/[1- {\sum}_{k}|V_{k}|^{2}
G_{0,\sigma} ({\bf k},\omega) D_{0,\sigma}(\omega)]
\end{equation}
Accordingly, the density of states for local electrons in the ballistic grains
is:
\begin{equation}
   \rho_{d,\sigma}(\omega)
   =\frac{1}{\pi} \frac{\Delta(\omega)}{D^{-2}_{0,\sigma}(\omega)+
    \Delta^{2}(\omega)}
\end{equation}
where the Friedel's halfwidth of the local electron [5] is:
\begin{equation}
    \Delta(\omega)=\pi \sum_{k}|V_{k}|^{2} \delta(\omega-
\epsilon_{k\sigma}+E_{F})
\end{equation}
where $\delta(\cdots)$ is the Dirac $\delta$-function.
In the present work, we consider the situation at zero temperature. Therefore,
the local electron occupation, ${\it n_{d}}$, and the LMM, ${\it m_{d}}$,
in the impurity site can be obtained through the preceding density of states
by the spectral theorem:
\begin{equation}
    n_{d}, m_{d} = \int^{E_{F}} d \omega \sum_{\sigma}(\sigma^{2},\sigma)
\rho_{d\sigma}(\omega) |_{B \rightarrow 0}
\end{equation}
in the absence of magnetic field.

   The contribution of the spin susceptibility comes from
the local electrons is obtained correspondingly. From
Eqs.(7) and (9), one can get the expression of the spin susceptibility from
local electrons at zero temperature:
\begin{equation}
    \chi_{d} = \frac{\partial m_{d}}{\partial B} |_{B\rightarrow 0}
=\int^{E_{F}} d\omega [\frac{\partial \rho_{d\uparrow}(\omega)}{\partial B} -
\frac{\partial \rho_{d\downarrow}(\omega)}{\partial B} ] |_{B \rightarrow 0}
\end{equation}
Thus the contribution of the local electrons to the spin susceptibility can be
obtained from the local density of states. In bulks,
the condition of the LMM formation is discussed in terms of the
instability of local spin susceptibility $^{9}$, later we will show
that it is difficult to do that for nanoscale metallic grains due to the
broken of transition invariance and the level fluctuation.  \\

\noindent {\bf 3.  Results and Discussions } \\

    With the above preparation, the effect of the level statistics on the local
magnetism in nanoscale metallic grains is considered and the
influence of the odd-even parity of conduction electrons on the LMM formation,
the critical condition for the LMM formation and the size effect of the local
spin susceptibility are discussed as the following. \\

\noindent {\bf 3.1  Local Magnetic Moment in Metallic Grains }

  According to  Eq.(9),  the local occupation and the LMM in impurity site at
zero temperature are obtained through the following expressions:
\begin{equation}
 {n_{d}, m_{d}} = \frac{1}{\pi} \int^{E_{F}} dw \sum_{\sigma}
({\sigma^{2}, \sigma}) \frac{\Delta_{\sigma}(w)}{
 (D_{0,\sigma}(\omega))^{-2} +\Delta_{\sigma}^{2}(w)}
\end{equation}
where
\begin{equation}
\Delta_{\sigma}(\omega)
=\pi V^{2}\sum_{k} {\delta}(\omega-\epsilon_{k\sigma}+E_{F})
\end{equation}
The main difference between the ultrasmall metallic grain
and the bulk lies in the summation over the wave-vector  ${\bf K}$ in Eq.(12).
In bulk metal, the summation over  ${\bf K}$ is quasi-continuous and can be
become of integration over ${\bf K}$, the density of states $\rho(0)$
of conduction electrons near the Fermi energy
is almost a constant, the hybridizing width
of the local electron is then: $\Delta_{bulk}=\pi \rho(0) V^{2}$. However, in
nanoscale metallic grains, the summation over ${\bf K}$ is discrete, so the
hybridizing width of the local electron is:
\begin{equation}
 \Delta_{\sigma}(\omega)
=\pi V^{2} \sum_{i} {\delta}(\omega-\epsilon_{i\sigma}+E_{F})
\end{equation}

As an approximation, $|V_{i}|$ is taken to be independent of the level
index. The average of this equation is taken over the disorder so that
all the statistical information about the level spacing contains in the
average over the summation of delta function. This can be related to the
two-level correlation and the nearest-neighbour level distribution
functions of the unitary ensemble, since the present system does not
preserve the time-reverse
invariance in the presence of spin-flipped scattering $^{10,11}$.

    In nanoscale metallic grains,
the eigenvalue energies of the conduction electrons are not
equi-spacing levels as suggested by Kubo $^{12}$ and Gorkov et al. $^{13}$,
rather, according to Wigner's random matrix theory $^{10,11}$, the energy level
obeys certain distributions for the nearest-neighbour level correlation and
the two level correlation, the distribution
function of the nearest-neighbour levels of the similar ensemble is:
\begin{equation}
   P_{n}(\epsilon)=c_{n}(\frac{\pi \epsilon}{\delta_{L}})^{n}
e^{-\pi\epsilon^{2}/(4{\delta_{L}}^{2})}
\end{equation}
where $c_{n}$ is the normalized coefficient, n is the ensemble symmetry
parameter, here \(n=2\), which corresponding to the unitary ensemble in the
presence of spin scattering in the systems containing magnetic impurity and in
magnetic field, and $c_{n}=\pi/2$;
P($\epsilon$)d$\epsilon$  represents the possibility of two nearest-neighbour
levels of the conduction electrons with separation $\epsilon$ near the Fermi
energy. And the two-level correlation function of the unitary ensemble
between levels $\epsilon$ and  $\epsilon^{'}$  is given by $^{13}$:
\begin{equation}
    R_{n}(\epsilon-\epsilon^{'})=1-Sin^{2}[\frac{\pi (\epsilon-\epsilon^{'})}
{\delta_{L}}]/[\frac{\pi (\epsilon-\epsilon^{'})}{\delta_{L}}]^{2}
\end{equation}
where $\epsilon$ and $\epsilon^{'}$ are two levels with arbitrary spacing in
the ensemble.
According to these two correlation functions, one can explore the effect
of the level statistics on the physical quantities of the system, while as,
these properties depend on the odd- or the even-parity of the
conduction-electron reservoir in the systems. Let
us consider the odd-parity case first.

   In the case of odd-parity, it is assumed that the chemical potential
coincides with the highest occupied level, $\epsilon_{1\sigma}=E_{F}$, which is
half-filled, therefore,
\begin{equation}
 \Delta_{odd}(\omega)
=\pi V^{2}\sum_{i} <{\delta}(\omega-
\epsilon_{i\sigma}+\epsilon_{1\sigma})>
\end{equation}
To a good approximation, one has $^{14}$:
\begin{equation}
 \delta_{L} \sum_{i} <{\delta}(\omega-\epsilon_{i}+\epsilon_{1})>
\approx R_{n}(\omega)
\end{equation}
so the averaged hybridizing width for odd-parity grains can be expressed as:
\begin{equation}
 \Delta_{odd}(\omega) = \frac{\pi V^{2}}{\delta_{L}} R(\omega)
\end{equation}
Substituting Eqs.(16-18) into Eq.(7), one could obtain the density of states
for local electrons in ultrasmall metallic grain with odd conduction electrons.

   Next we consider the case of even-parity. In this situation, the chemical
potential doesn't coincides with the highest occupied level $\epsilon_{0}$.
Assuming that the chemical potential lies in the half-way between the highest
occupied level, $\epsilon_{1}$, and the first empty level, $\epsilon_{0}$, or
$\mu=(\epsilon_{1}+\epsilon_{0})/2$, so the averaged hybridizing width of
the local electrons for even-parity grains can be written as:
\begin{eqnarray}
 \Delta_{even}(\omega)
=\pi V^{2}\sum_{i} <{\delta}[\omega-(\epsilon_{i\sigma}- \epsilon_{1\sigma})
+\frac{\epsilon_{0\sigma}-\epsilon_{1\sigma}}{2} ]>  \nonumber  \\
  =\frac{\pi V^{2}}{\delta_{L}} P_{n}(\omega)
+ \frac{\pi V^{2}}{\delta_{L}^{2}} \int^{E_{F}}_{0} d\epsilon
R_{n}(\epsilon) P_{n}[2(\epsilon-\omega)]
\end{eqnarray}
The first term in the second line in Eq.(19) comes from the summation over
energy index {\it i}
when $\epsilon_{i}$=$\epsilon_{1}$, the definitions of two-level correlation
function, $P_{n}(x)$, and the distribution function of the
nearest-neighbour, $R_{n}(x)$, are given in Eqs.(14) and (15),
respectively. Since the averaged hybridizing width of the local electron
with the conduction electrons depends on the energy $\omega$ through a
complicated form, the integrals in Eqs.(18) and (19) can not be performed
analytically, numerical calculation is thus
performed for some typical systems and the results are shown in the following.

    It is found that some features of local electronic states and the LMM in
nanoscale grain are similar to that
of bulk. The formation of the Friedel's resonance states between the impurity
electron and the conduction electrons through the hybridizing interaction and
the splitting of the Friedel's resonance states with up and down spins give
rise to the LMM when the Coulomb interaction exceeds a critical value. With the
increase of hybridizing interaction $V_{k}$,  the impurity electron becomes
more and more delocalized so that  the mean occupation at impurity site and the
LMM becomes small.

However, one striking difference of the LMM in nanoscale grain from that
in bulk is that the critical Coulomb interaction $U_{c}$ for the occurrence
of LMM in the grains with odd conduction electrons is much larger than that
with even conduction electrons.
The Coulomb-repulsion dependence of the LMM in metallic grain with
odd- and even-parity is shown in Fig.1. The LMM of nanoscale metallic grains
strongly depending on the parity originates from the level statistics, or the
position of the chemical
potential with respect to the occupied energy levels, especially to the highest
occupied level. In the odd-parity  grains, the broaden halfwidth of the local
electron, $\Delta_{odd}$, is relevant to the two-level correlation function
$R_{n}$, which is only affected by a few levels close to the chemical
potential. In
the even grains, the halfwidth, $\Delta_{even}$, depends on not only the two-
level correlation function but also the nearest-neighbour level distribution
function $P_{n}$ due to strong level repulsion. Because of the strong level
repulsion, only several levels below the
chemical potential contribute to the density of states of the local electrons
for the latter. For the systems with the same interaction parameters, the mean
occupation of local
electron and the LMM in odd-parity grains are larger than those in even-parity
grains. This can be seen clearly in Fig.1. Therefore, unsmooth variation
of the magnetic moment in nanoscale grain is expected
when the electron number in the grain increases gradually.

 For large Coulomb interaction, the differences of the local occupation
and the LMM for odd- and even-grains become small. Curve 1 and 2 , Curve 3
and 4 in Fig.1 approach the same values
for the fact that when U becomes large, the two levels of the resonance states
are separated so large that the electron occupies the lower state.
the local electronic state, hence the local moment, is almost not affected
by the environmental electric field, so the local electronic states and the
LMM become independent of the parity gradually.

   Another one important effect of the level statistics is that the on-site
Coulomb interaction necessary for the spin-splitting of the resonance states
and the occurrence of the LMM is different for the odd-parity  grain and for
the even-parity grains. It strongly depends on the grain size.
The general feature of the size-dependence of the critical Coulomb-strength
for the occurrence of the LMM is shown in Fig.2. One finds that the critical
on-site Coulomb interaction, $U_{c}$, increases when the grain size becomes
large, and the $U_{c}$ in odd-parity grain  is much smaller than that in
even-parity grain.

  As we all know, the condition for the occurrence of LMM in the Anderson model
in bulk is  $\rho_{d}(0)U_{c}=1$ when the impurity level lies below the Fermi
energy. However this condition no longer holds for nanoscale metallic
grains. The critical value  $\rho_{d}(0)U_{c}$ behaves strong size-dependence
and relates to the parity of grains. The numerical result is shown in Fig.3.
The parameters are the same as those in Fig.2. We find that the critical value
is much smaller than unity and increases with the size of grain. This can
be explained in terms of the strong repulsion of the energy
levels and the strong localization of the impurity electronic states in finite
size system. The reason of the critical value, $\rho(0)U_{c}$, in even grains
is generally larger than that in odd grains is the same as  that of $U_{c}$.

   The size-dependences of the mean local occupation and the LMM
are shown in Fig.4. The main feature that the mean occupation and the LMM
decrease as the increase of size is very similar to that of clusters so that
the finite size effect is obvious. For the same metallic systems, it is more
difficult for the
occurrence of the LMM in the grains with even-parity than that with odd-parity.
It also can be attributed to the fact that the resonance halfwidth of the
former is larger than that of the latter.
Delocalization character of the impurity electron becomes dominant
when the grain size becomes large. The  local
occupation and the LMM approach constants (the bulk values) gradually
are possibly expected when the
size becomes very large. It was not found in the present
size range (N $\leq$ 3000), however. Similar situation also happens for the
critical Coulomb interaction in Fig.2. It is believed that much large size
grain should be considered for the present purpose. \\

\noindent {\bf 3.2 Local Spin Susceptibility in Metallic Grains } \\

    We now turn to the local spin susceptibility of nanoscale metallic grains
at zero temperature. It is
reasonable to assume that the hybridizing width of the local electrons
is independent of the magnetic field, so the expression of the local spin
susceptibility is rewritten as:
\begin{equation}
    \chi_{d} = \frac{\pi}{2} \sum_{\sigma} \int^{E_{F}} d\omega
  \frac{{\Delta}(\omega)(\omega-\epsilon_{d}- U<n_{\bar{\sigma}}>)}{(\omega-
  \epsilon_{d} -U <n_{\bar{\sigma}}>)^{2}+{\Delta}^{2}(\omega)}
\end{equation}
where the local spin susceptibility is measured in units of the square of the
product of $g$ factor and the Bohr magneton $\mu_{B}$, i.e. $(g\mu_{B})^{2}$.
The behavior of the local spin susceptibility for the grains with odd- and with
even-conduction electrons in accordance with the hybridizing width
$\Delta_{odd}$ or $\Delta_{even}$ is then examined. The numerical
results are shown in Fig.5 and Fig.6. \\

 It can be seen in Fig.5 that the reduced local spin susceptibility,
$\chi_{d}$, depends on the on-site Coulomb-interaction of the local electrons.
As a comparison, the dependence of the LMM on size is also shown in this figure.
Before the formation of LMM, $\chi_{d}$ decreases as
the increase of the Coulomb interaction  U .  After the LMM occurs, $\chi_{d}$
increases with U. Detail calculation shows that
the local susceptibility exhibits a finite discontinuity at the
critical Coulomb strength for the occurrence of LMM.
The reason of the decrease of  $\chi_{d}$ below $U_{c}$ and the increase
$\chi_{d}$ above $U_{c}$ lies in the different response of the local electron
to external magnetic field. For smaller U than $U_{c}$, there is no
spin-splitting for local states, $\chi_{d}$ is positive, the local electron
exhibits paramagnetic character. When U approaches $U_{c}$, the LMM occurs
and $\chi_{d}$ varies crucially, but doesn't diverge as in bulk due to the
finite size effect. It exhibits a finite discontinuity at $U_{c}$.

Fig.6 shows the evolution of the local spin susceptibility with the increase of
grain size. For the grain with small size, $\chi_{d}$ exhibits strong
diamagnetic character, in which the LMM occurs. For
the grain with large size, \( \chi_{d} \) appears weak paramagnetic
character since the LMM
disappears. When the size becomes very large, the effect of odd-even parity
becomes of unimportant.

  In an earlier study $^{15}$, we found that if only the nearest-neighbour
level
distribution function was taken into account, the local magnetism would be much
weaker than that in the present case. This can be attributed to the fact that
when the
two level correlation is absence, the impurity electron is more itinerant.
The presence of the two level correlation will enhance the localization
character of impurity electron. Also, the level statistic affects some other
physical properties of metallic grains $^{16}$.  Also the present method can
be generalized to the
nanoscale metallic grains in the diffusive regime, in which $1/\tau$ is
finite and the fluctuations of LMM and local spin susceptibility are
expected to be large.\\

\noindent {\bf 4. CONCLUSION} \\

     The quantum correction arising from the level statistics effect to the
local magnetic moments and the local spin susceptibility in the nanoscale
ballistic metallic grains are important. The present study suggests that the
condition for the occurrence of the LMM still exists with the quality much
smaller than that of bulk and depends on the size and parity of nanoscale
grains.\\

 {\it Acknowledgment}: One of authors, Zou, thanks Zeng Zi for reading the
manuscript, and thanks Yu Lu and H. A. Cerdeira
and the hospitality of the International Centre for Theoretical Physics (ICTP)
in Trieste, Italy; This work is partially supported by the Grant of NSF of
China and the Panden Project. The authors are grateful to the Physical
Society of Japan for financial support in publication.

\newpage
\large
\begin{center}
REFERENCES
\end{center}

\begin{enumerate}

\item G. M. Patsor, R. Hirsch and B. Muhlschlegel,~~ Phys. Rev. Lett. {\bf 72},
      3879 (1994);~~ G. M. Pastor, J. Dorantes-Davila and K.H.Bennemann,~~
      Phys. Rev. {\bf B40}, 7642 (1989)
\item B. V. Reddy, S. N.Khanna  and B. I. Dunlap,~~ Phys. Rev. Lett. {\bf 70},
      3323 (1993).  S. N. Khanna and S. Linderoth, ~~ Phys. Rev. Lett. {\bf
      67}, 742 (1991);  J. Magn. Magn. Mater. {\bf 104-107}, 1574 (1992).
\item H. Yoshioka {\it J. Phys. Soc. Jpn},{\bf 63}, 405, (1994).
\item H. Mathur, M. Gokcedag and A. D. Stone,
      {\it Phys. Rev. Lett.} {\bf 74} 1855 (1995).
\item P. W. Anderson, ~~ Phys. Rev. {\bf 124}, 41 (1961);~~ Rev. Mod. Phys.
      {\bf B50}, 191 (1978), and some references therein.
\item B. L. Altshuler and B. I. Shklovskii, Sov. Phys. JETP, {\it 64}, 127
      (1986).
\item Y. Imry, Transport Phenomena in Mesoscopic Systems, ed. by H. Fukuyama
      and T. Ando, (Springer-Verlag Berlin Heidelberg, 1992), P205.
\item T. M. Rice, {\it Metal-Insulator Transitions}, Troisieme Cycle de la
      physique, Semmester d'hiver, 1983-1984. Zurich.
\item S. Doniach and E. H. Sondheimer, {\it Green's Functions for Solid State
      Physicists}, (W. A. Benjamin, Inc. MA, 1974).
\item E. P. Wigner, Ann. Math. {\bf 53}, 36 (1951); {\bf 62}, 548 (1955);
      {\bf 65}, 203 (1957); {\bf 67}, 325 (1958).
\item K. B. Efetov, Adv. Phys. {\bf 32}, 53 (1983).
\item R. Kubo, J. Phys. Soc. Jpn., {\bf 17}, 975 (1962).
\item L. P. Gorkov and G. M. Eliashberg, Zh. Eksp. Teor. Fiz., {\bf 43}, 1407
      (1965).
\item R. A. Smith and V. Ambegaokar, {\it Phys. Rev. Lett.}, {\bf 77}, 4962
      (1996).
\item Liang-Jian Zou, Zheng Qing-Qi and X. G. Gong, {\it Chin. Phys. Lett.}
      {\bf 14}, 383 (1997).
\item A. Cerdeira, B. Kramer and G. Schon, eds. Quantum Dynamics of Submicron
      Structure, NATO ASI series, (Kluwer Academic Publishers, Dordrecht,
      The Netherlands, 1995).

\end{enumerate}

\newpage
\large

\begin{center}
Figures Captions
\end{center}

\noindent  Fig.1. Coulomb-repulsion dependence of the mean occupation (Curve
           1 and 2) and the local magnetic moment (Curve 3 and 4) in the 
           ballistic metallic grains, parameters
           $\epsilon_{d}$=-0.5, $E_{F}$=3.0, V=0.1. Energy is in units of eV,
           magnetic moment is in units of $\mu_{B}$.   \\

\noindent  Fig.2. Size-dependence of the critical Coulomb-interaction for the
           occurrence of the local magnetic moment. Here
           $\epsilon_{d}$=-0.5, $E_{F}$=8.0, V=0.1. Energy is in units of eV. \\

\noindent  Fig.3. Dependence of the critical value, $\rho_{d}(0)U_{c}$, for the
           occurrence of local magnetic moment on size. Parameters are the same
           as those in Fig.2.  \\

\noindent  Fig.4. Size-dependence of the mean occupation (Curve 1 and 2) and the 
           local magnetic moment for odd-parity and even-parity metallic grains.
           $\epsilon_{d}$=-2.5, $E_{F}$=8.0, U=6.0, V=0.1. Energy is in
           units ofeV. magnetic moment in units of $\mu_{B}$.\\

\noindent  Fig.5. Dependence of the local spin susceptibility (Curve 2 and 4)
           and the local magnetic moment (Curve 1 and 3) on the Coulomb
           interaction for metallic grains
           with odd (N=101) and even (N=100) conduction electrons.
           $\epsilon_{d}$=-0.5, $E_{F}$=3.0, V=0.1. Energy is in units of eV,
           Susceptibility is in units of $(g\mu_{B})^{2}$.   \\

\noindent  Fig.6 Dependence of local spin susceptibility and local
           magnetic moment on the size of metallic grains with odd
           and even conduction electrons.
           $\epsilon_{d}$=-0.5, $E_{F}$=5.0, U=6.0, V=0.1. Energy is in
           units of eV.

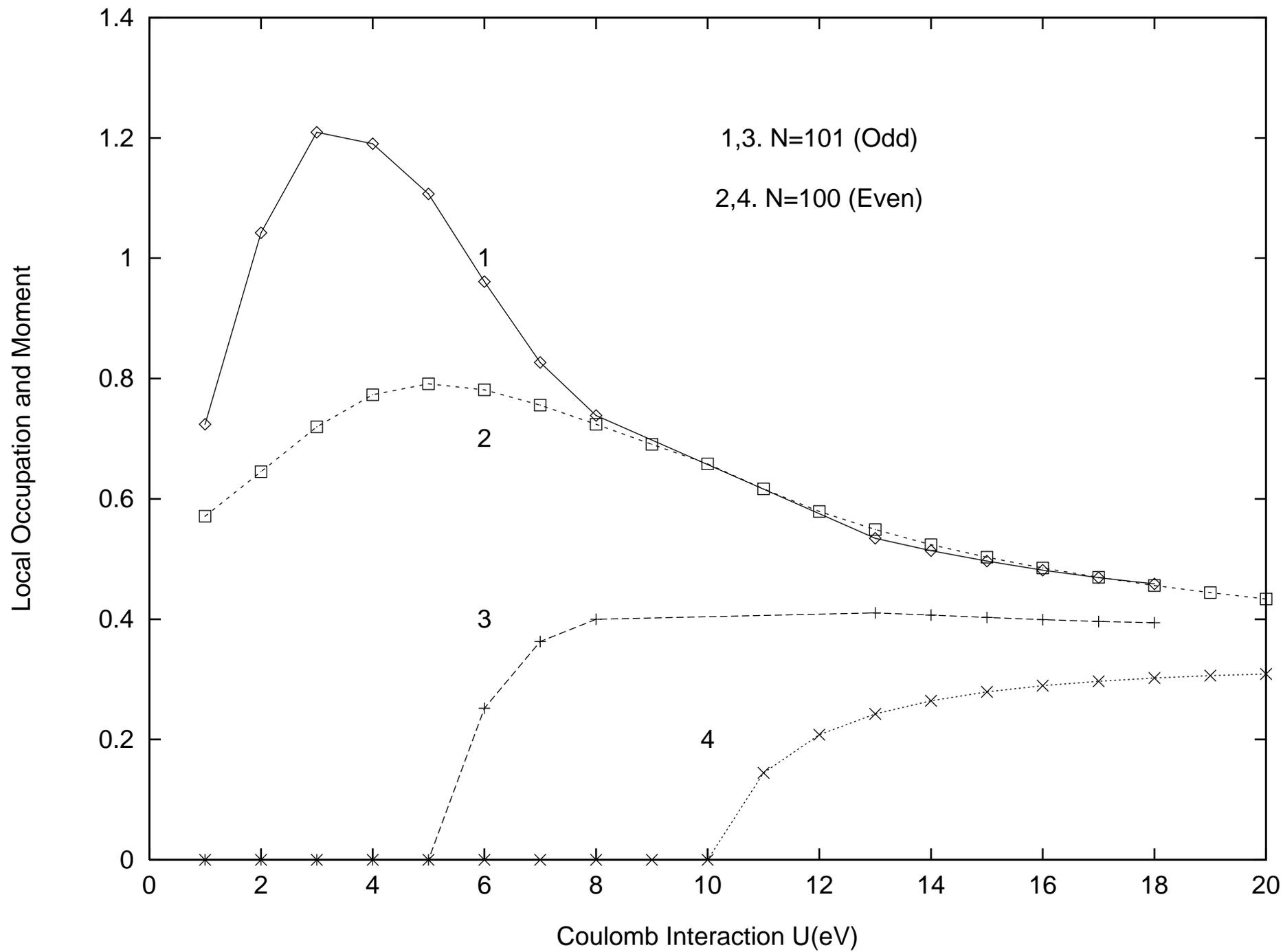
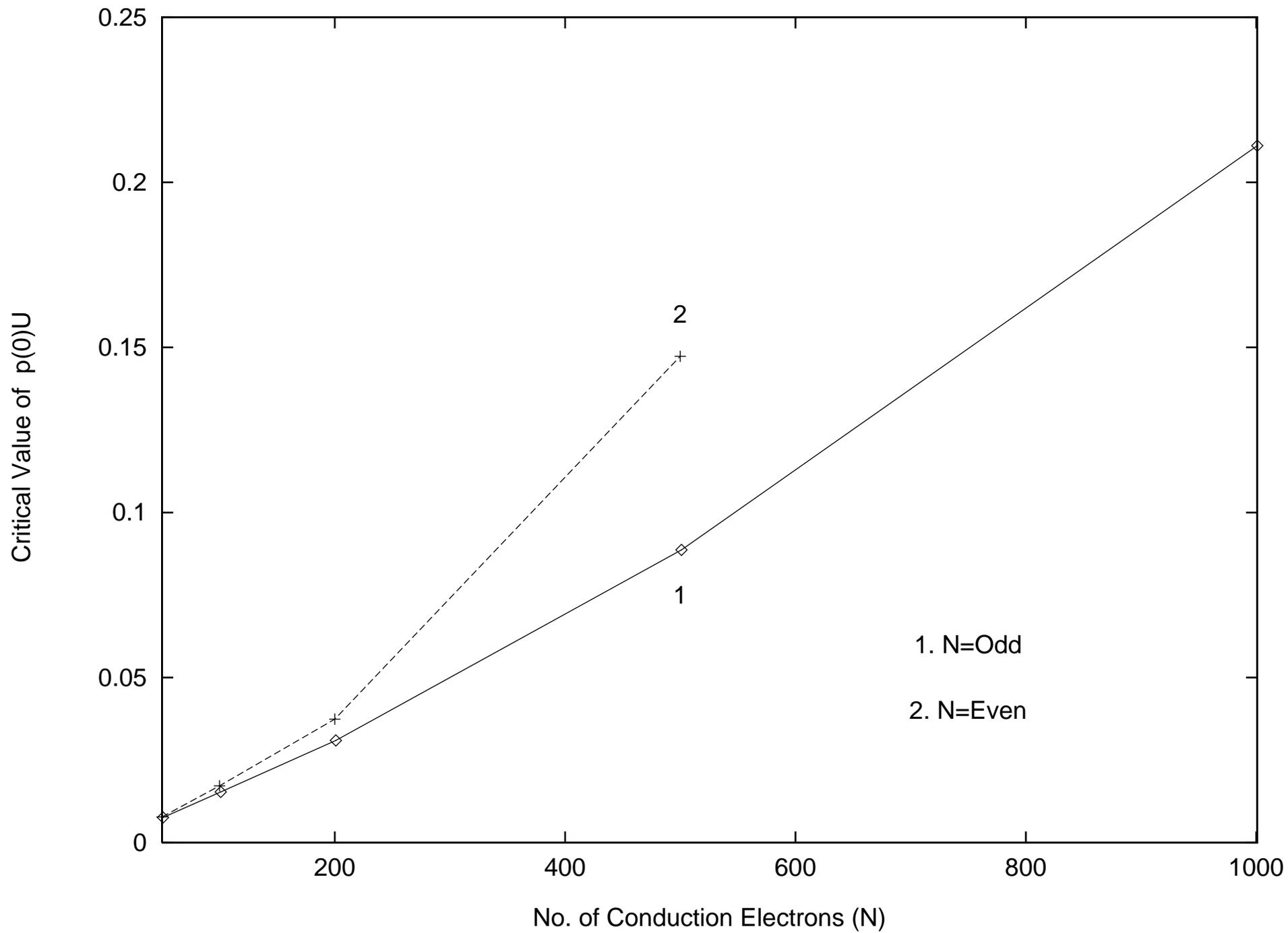
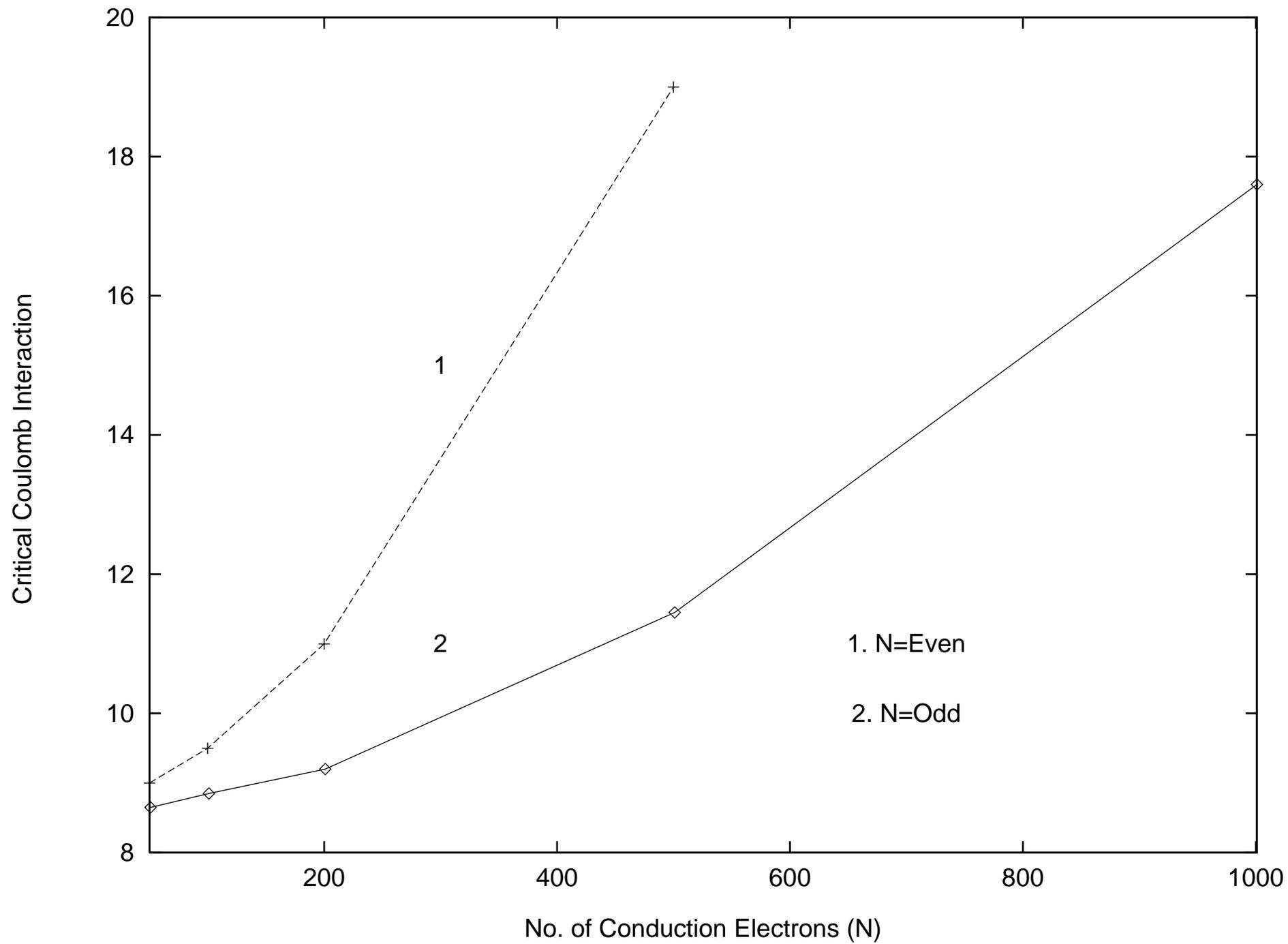
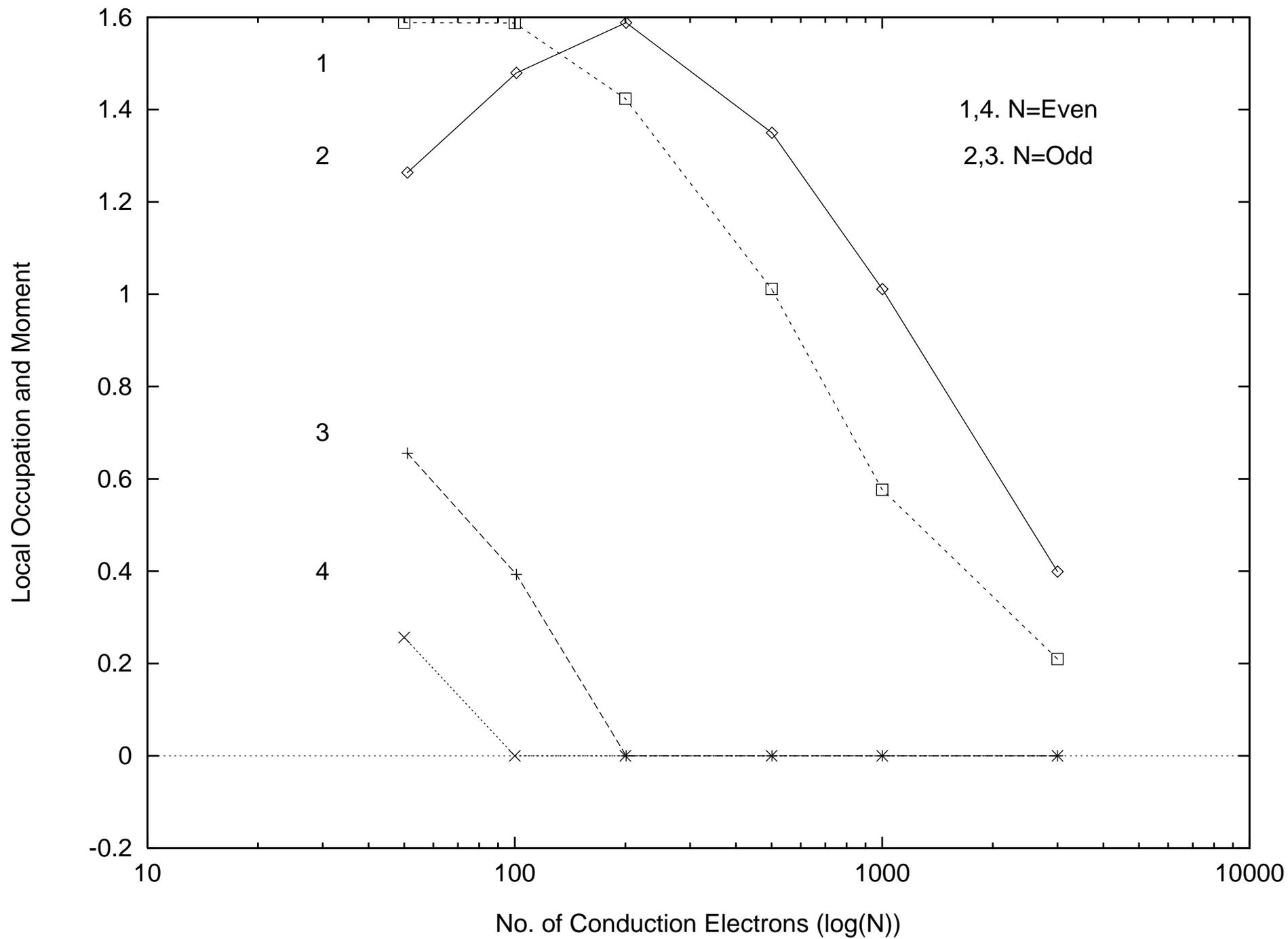
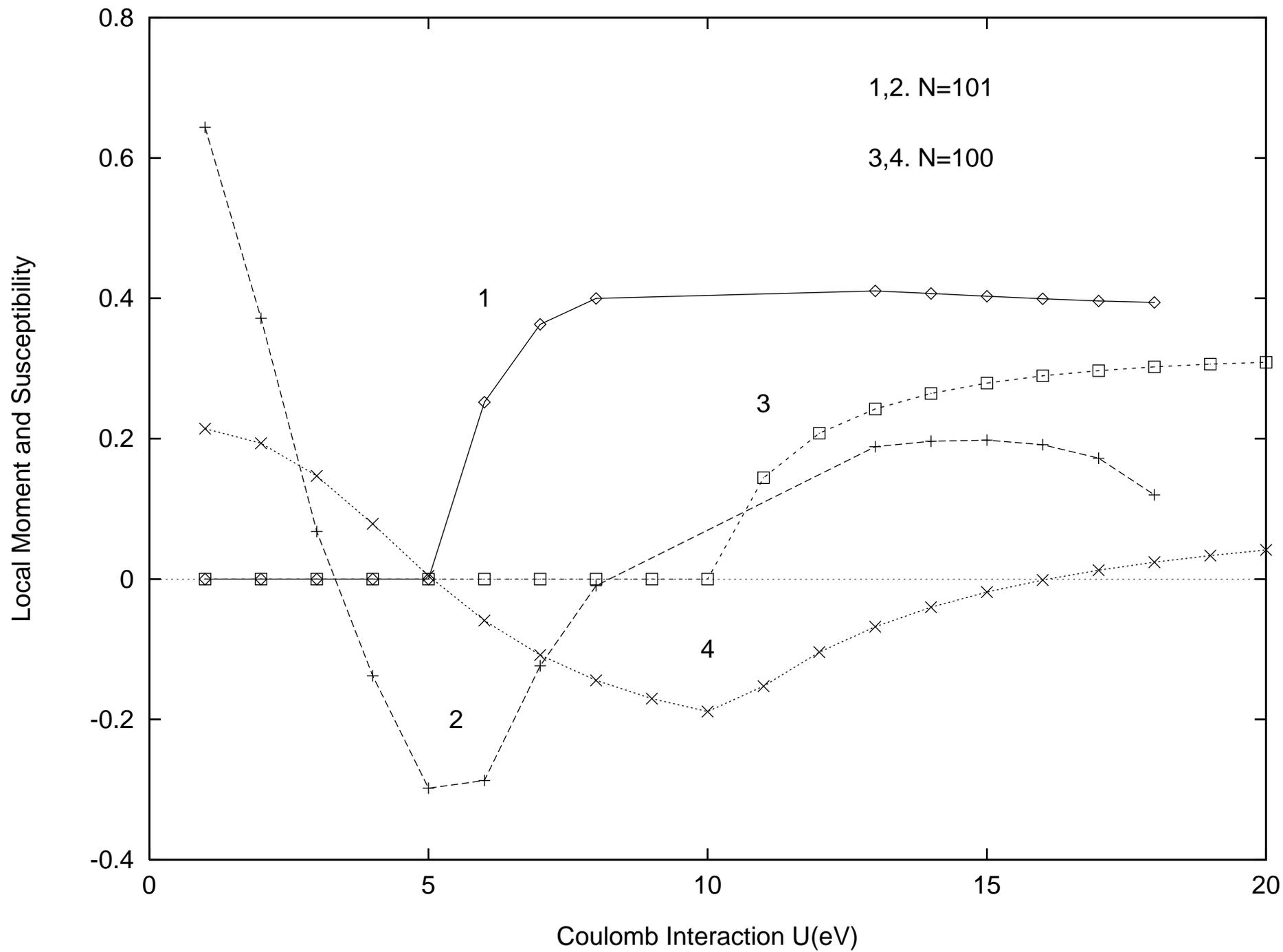
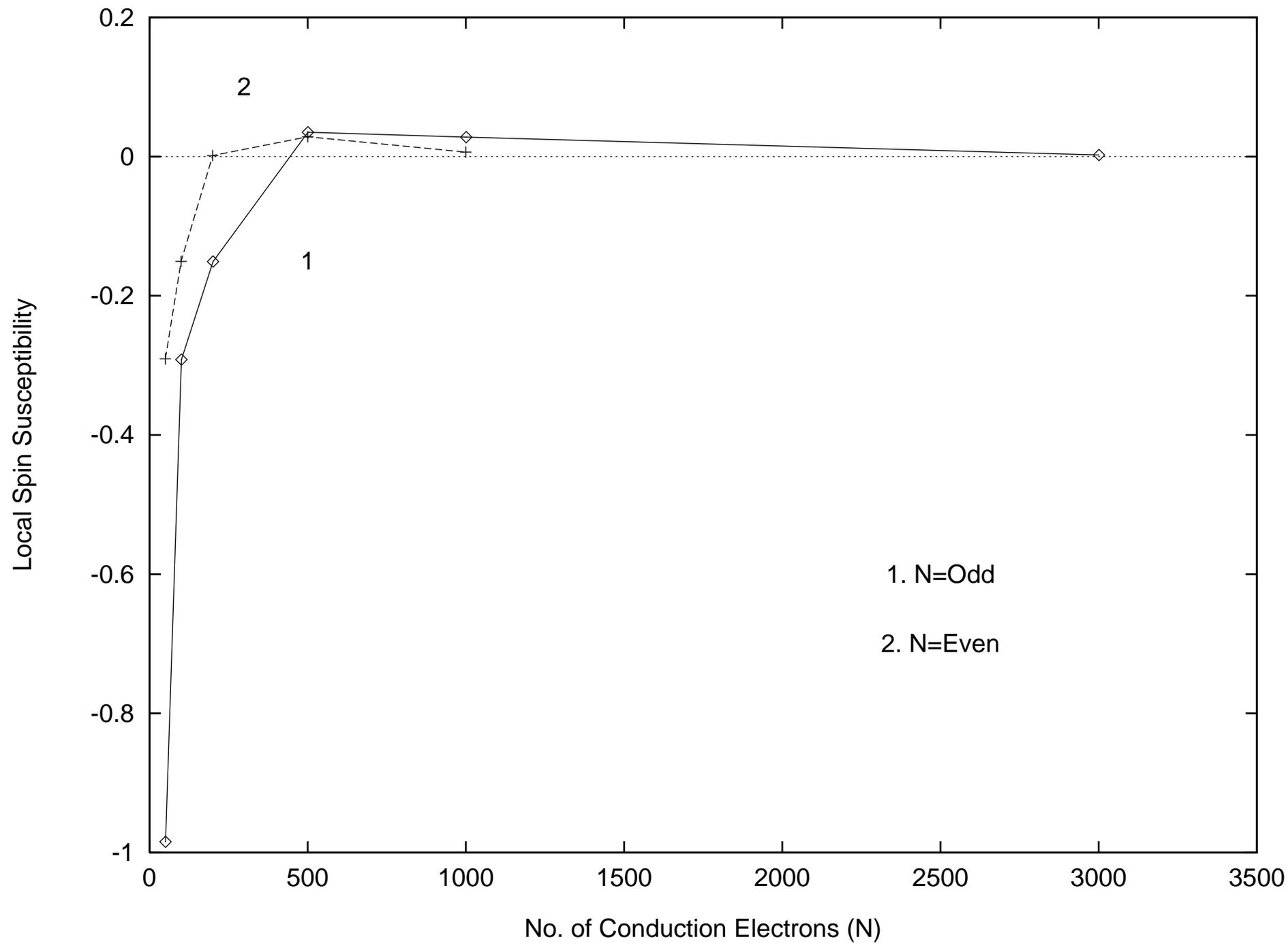
\end{document}